\documentclass[12pt]{article}
\usepackage{a4wide,epsfig}

\usepackage{slashed}

\newcommand{\agt}{\rlap{\lower 3.5 pt \hbox{$\mathchar \sim$}} \raise 1pt
 \hbox {$>$}}
\newcommand{\alt}{\rlap{\lower 3.5 pt \hbox{$\mathchar \sim$}} \raise 1pt
 \hbox {$<$}}

\catcode`@=11
\newcount\@tempcntc
\def\@citex[#1]#2{\if@filesw\immediate\write\@auxout{\string\citation{#2}}\fi
  \@tempcnta\z@\@tempcntb\m@ne\def\@citea{}\@cite{\@for\@citeb:=#2\do
    {\@ifundefined
       {b@\@citeb}{\@citeo\@tempcntb\m@ne\@citea\def\@citea{,}{\bf
?}\@warning
       {Citation `\@citeb' on page \thepage \space undefined}}%
    {\setbox\z@\hbox{\global\@tempcntc0\csname b@\@citeb\endcsname\relax}%
     \ifnum\@tempcntc=\z@ \@citeo\@tempcntb\m@ne
       \@citea\def\@citea{,}\hbox{\csname b@\@citeb\endcsname}%
     \else
      \advance\@tempcntb\@ne
      \ifnum\@tempcntb=\@tempcntc
      \else\advance\@tempcntb\m@ne\@citeo
      \@tempcnta\@tempcntc\@tempcntb\@tempcntc\fi\fi}}\@citeo}{#1}}
\def\@citeo{\ifnum\@tempcnta>\@tempcntb\else\@citea\def\@citea{,}%
  \ifnum\@tempcnta=\@tempcntb\the\@tempcnta\else
   {\advance\@tempcnta\@ne\ifnum\@tempcnta=\@tempcntb \else
\def\@citea{--}\fi
    \advance\@tempcnta\m@ne\the\@tempcnta\@citea\the\@tempcntb}\fi\fi}
\catcode`@=12

\begin{document}

\title{
\vskip-3cm{\baselineskip14pt
\centerline{\normalsize DESY 06-235\hfill ISSN 0418-9833}
\centerline{\normalsize hep-ph/0612184\hfill}
\centerline{\normalsize December 2006\hfill}}
\vskip1.5cm
Two-loop virtual top-quark effect on Higgs-boson decay to bottom quarks}

\author{Mathias Butensch\"on, Frank Fugel, Bernd A. Kniehl\\
{\normalsize II. Institut f\"ur Theoretische Physik, Universit\"at Hamburg,}\\
{\normalsize Luruper Chaussee 149, 22761 Hamburg, Germany}
}
\date{}

\maketitle

\begin{abstract}
In most of the mass range encompassed by the limits from the direct search and
the electroweak precision tests, the Higgs boson of the standard model
preferably decays to bottom quarks.
We present, in analytic form, the dominant two-loop electroweak correction, of
$\mathcal{O}(G_F^2m_t^4)$, to the partial width of this decay.
It amplifies the familiar enhancement due to the ${\cal O}(G_Fm_t^2)$ one-loop
correction by about $+16\%$ and thus more than compensates the screening by
about $-8\%$ through strong-interaction effects of order
${\cal O}(\alpha_sG_Fm_t^2)$.

\medskip

\end{abstract}

\newpage

The standard model (SM) of elementary-particle physics, whose fermion and
gauge sectors have been impressively confirmed by an enormous wealth of
experimental data, predicts the existence of a last undiscovered fundamental
particle, the Higgs boson $H$, whose mass $M_H$ is a free parameter of the
theory.
The direct search for the Higgs boson at the CERN Large Electron-Positron
Collider LEP~2 led to a lower bound of $M_H>114$~GeV at 95\% confidence
level (CL) \cite{Barate:2003sz}.
On the other hand, high-precision measurements, especially at LEP and the SLAC
Linear Collider SLC, were sensitive to the Higgs-boson mass via electroweak
radiative corrections, yielding to the value
$M_H=\left(85^{+39}_{-28}\right)$~GeV together with an upper limit of
$M_H<166$~GeV at 95\% CL \cite{LEPEWWG}. 
The vacuum-stability and triviality bounds suggest that
$130\alt M_H\alt 180$~GeV if the SM is valid up to the grand-unification scale
(for a review, see Ref.~\cite{Kniehl:2001jy}).
If the Higgs mechanism of spontaneous symmetry breaking, as implemented in the
SM, is realized in nature, then we are now being on the eve of a
groundbreaking discovery, to be made at the CERN Large Hadron Collider (LHC),
which will go into operation in a just few months from now.
After finding a new scalar particle, the burning question will be whether it
is in fact the Higgs boson of the SM, or lives in some extended Higgs sector.
Therefore, it is indispensable to know the SM predictions for the production
and decay rates of the SM Higgs boson with high precision.
Its decay to a bottom-quark pair, $H\to b\overline{b}$, is of paramount
interest, as it is by far the dominant decay channel for $M_H\alt140$~GeV
(see, {\it e.g.}, Ref.~\cite{Kniehl:1993ay}).
On the other hand, the inverse process, $b\overline{b}\to H$, was identified 
to be a crucial hadroproduction mechanism, appreciably enhancing the yield due
to gluon fusion \cite{Maltoni:2003pn}.
Precise knowledge of the bottom Yukawa coupling is also requisite for reliable
predictions of associated hadroproduction of Higgs bosons and bottom quarks
\cite{Dawson:2003kb}.

The purpose of this Letter is to fill a long-standing gap in our knowledge of
the quantum corrections to the partial width $\Gamma_b$ of the
$H\to b\overline{b}$ decay, by providing, in analytic form, the dominant
two-loop electroweak correction, of ${\cal O}(G_F^2m_t^4)$, where $G_F$ is
Fermi's constant and $m_t$ is the top-quark mass.
This correction also applies to the cross section of $b\overline{b}\to H$.
Surprisingly, it turns out to be more than twice as large as the
${\cal O}(\alpha_sG_Fm_t^2)$ one, which is formally enhanced by one power of
the strong-coupling constant $\alpha_s$.
In the discussion of virtual top-quark effects, it is useful to distinguish
between universal corrections, which are independent of the produced fermion
flavor, and non-universal corrections, which are specific for the
$H\to b\overline{b}$ decay because bottom is the weak-isospin partner of top.
Here, we have to consider both types.

Prior to going into details with our calculation, we briefly review the
current status of the radiative corrections to $\Gamma_b$ in the intermediate
mass range, defined by $M_W<M_H<2M_W$.
As for effects arising solely from quantum chromodynamics (QCD), the full
$m_b$ dependence is known in ${\cal O}(\alpha_s)$ \cite{Braaten:1980yq}.
In ${\cal O}(\alpha_s^2)$, the leading \cite{Gorishnii:1991zr} and
next-to-leading \cite{Surguladze:1994gc} terms of the expansion in
$m_b^2/M_H^2$ of the Feynman diagrams without top quarks are available.
Those involving top quarks either contain gluon self-energy insertions or
represent cuts through three-loop double-triangle diagrams;
the former contribution is exactly known \cite{Kniehl:1994vq}, while the
four leading terms of the expansion in $M_H^2/m_t^2$ are known in the latter
case \cite{Chetyrkin:1995pd}.
In ${\cal O}(\alpha_s^3)$, the diagrams containing only light degrees of
freedom were evaluated directly \cite{Chetyrkin:1996sr}, while those involving
the top quark were treated in the framework of an appropriate effective field
theory \cite{Chetyrkin:1997vj}.
As for purely electroweak corrections, the one-loop result is completely known
\cite{Kniehl:1991}.
At two loops, the dominant universal correction, of ${\cal O}(G_F^2m_t^4)$,
was already studied in Ref.~\cite{Djouadi:1997rj}, while the non-universal one
is considered here for the first time.
As for mixed corrections, the universal \cite{Kniehl:1994ph} and non-universal
\cite{Kniehl:1994ju} ${\cal O}(\alpha_sG_Fm_t^2)$ terms at two loops and the
universal \cite{Kniehl:1995br} and non-universal \cite{Chetyrkin:1996ke}
${\cal O}(\alpha_s^2G_Fm_t^2)$ terms at three loops are available.

We now outline the course of our calculation and exhibit the structure of our
results.
Full details will be presented in a forthcoming communication \cite{long}.
For convenience, we work in 't~Hooft-Feynman gauge.
As usual, we extract the ultraviolet divergences by means of dimensional
regularization, with $D=4-2\epsilon$ space-time dimensions and 't~Hooft mass
scale $\mu$.
We do not encounter ambiguities related to the treatment of $\gamma_5$ in $D$
dimensions and are thus entitled to use the anti-commuting definition.
We adopt Sirlin's formulation of the electroweak on-shell renormalization
scheme \cite{Sirlin:1980nh}, which uses $G_F$ and the physical particle masses
as basic parameters.
We take the Cabibbo-Kobayashi-Maskawa quark mixing matrix to be unity, which
is well justified because the third quark generation is, to good
approximation, decoupled from the first two \cite{pdg}.
For convenience, we renormalize the Higgs sector by introducing counterterm
vertices involving tadpole and Higgs-boson mass counterterms, $\delta t$ and 
$\delta M_H$, respectively \cite{Denner:1991kt}.
Specifically, $\delta t$ is adjusted so that it exactly cancels the sum of all
one-particle-irreducible tadpole diagrams.

Detailed inspection reveals that, to the orders considered here, the amputated
matrix element of $H\to b\overline{b}$ exhibits the simple structure
\begin{equation}
\mathcal{A}=A+B\left(\slashed{p}-\slashed{\overline{p}}\right)\omega_-,
\end{equation}
where $\omega_\pm=(1\pm\gamma_5)/2$ are the helicity projection operators,
$p$ and $\overline{p}$ are the four-momenta of $b$ and $\overline{b}$,
respectively, and $A$ and $B$ are Lorentz scalars.
Including the wave-function renormalizations of the external particles and
employing the Dirac equation, we find the transition matrix element to be
\begin{equation}
\mathcal{T}=\sqrt{Z_H}\left(\sqrt{Z_{b,L}Z_{b,R}}A+m_bZ_{b,L}B\right)s,
\end{equation}
where $s=\overline{u}(p,r)v(\overline{p},\overline{r})$, with $r$ and
$\overline{r}$ being spin labels.
Owing to parity violation, the left- and right-handed components of the
bottom-quark field, $b_{L,R}=\omega_\mp b$, participate differently in the
electroweak interactions and thus receive different wave-function
renormalizations, $Z_{b,L/R}$.
At tree-level, we have $A^{(0)}=-m_b/v$ and $B^{(0)}=0$, where
$v=2^{-1/4}G_F^{-1/2}$ is the Higgs vacuum expectation value.
Here and in the following, superscripts enclosed in parentheses denote the
loop order.
In Sirlin's formulation of the electroweak on-shell scheme, where Fermi's
constant is introduced to the SM through a charged-current process, namely
muon decay, the SU(2) gauge coupling $g=2M_W/v$ does not receive power
corrections in $m_t$, so that \cite{Consoli:1989fg}
\begin{equation}
\frac{M_{W,0}}{v_0}=\frac{M_W}{v}
\end{equation}
to the orders considered here, which implies that the renormalization of $v$
is reduced to the one of $M_W$.
Here and in the following, bare quantities carry the subscript 0.
It hence follows that we need to perform a genuine two-loop renormalization of
$Z_H$, $m_b$, $Z_{b,L/R}$, and $M_W$, while a one-loop renormalization of
$M_H$ and $m_t$ is sufficient.
As usual, we denote the sums of all one-particle-irreducible $H$, $f$
($f=b,t$), and $W$ self-energy diagrams at four-momentum transfer $q$ as
$i\Sigma_H(q^2)$,
$i[\slashed{q}(\omega_-\Sigma_{f,L}(q^2)+\omega_+\Sigma_{f,R}(q^2))
+m_{f,0}\Sigma_{f,S}(q^2)]$, and
$-i[(g^{\mu\nu}-q^\mu q^\nu/q^2)\Sigma_{W,T}(q^2)
+(q^\mu q^\nu/q^2)\Sigma_{W,L}(q^2)]$, and
split the bare masses as $M_{H/W,0}^2=M_{H/W}^2+\delta M_{H/W}^2$ and
$m_{f,0}=m_f+\delta m_f$.
Imposing the on-shell renormalization conditions on the dressed propagators
then yields
\begin{eqnarray}
\delta M_H^2&=&\Sigma_{H}(M_H^2),
\nonumber\\
Z_H&=&\frac{1}{1+\Sigma_H^\prime(M_H^2)},
\nonumber\\
\frac{\delta m_f}{m_f}&=&\frac{1}{\sqrt{f(m_f^2)}}-1,
\nonumber\\
Z_{f,L/R}&=&\frac{1}{\left(1+\Sigma_{f,L/R}(m_f^2)\right)
\left(1-m_f^2\frac{f^\prime(m_f^2)}{f(m_f^2)}\right)},
\nonumber\\
\delta M_W^2&=&\Sigma_{W,T}(M_W^2),
\label{eq:ren}
\end{eqnarray}
where
\begin{equation}
f(q^2)=\frac{(1-\Sigma_{f,S}(q^2))^2}{(1+\Sigma_{f,L}(q^2))
(1+\Sigma_{f,R}(q^2))}.
\end{equation}
Relations that, to the order of our analysis, are equivalent to
Eq.~(\ref{eq:ren}) were found in Ref.~\cite{Faisst} using an alternative
approach.

Performing a loop expansion and eliminating all bare masses, we thus obtain
\begin{eqnarray}
\lefteqn{\frac{\mathcal{T}^{(0)}}{s}=A^{(0)},}
\nonumber\\
\lefteqn{\frac{\mathcal{T}^{(1)}}{s}=A^{(1)}+m_bB^{(1)}
+A^{(0)}\left(\delta_u^{(1)}+X^{(1)}\right),}
\nonumber\\
\lefteqn{\frac{\mathcal{T}^{(2)}}{s}=A^{(2)}+m_bB^{(2)}
+A^{(1)}X^{(1)}+m_bB^{(1)}\delta Z_{b,L}^{(1)}}
\nonumber\\
&&{}+\left(A^{(1)}+m_bB^{(1)}+A^{(0)}X^{(1)}\right)
\left[\delta_u^{(1)}
+2(1-\epsilon)\frac{\delta m_t^{(1)}}{m_t}
\right.
\nonumber\\
&&{}-\left.\frac{\delta M_W^{2(1)}}{M_W^2}\right]
+A^{(0)}\left[\delta_u^{(2)}+X^{(2)}
+\frac{1}{2}\,\frac{\delta m_b^{(1)}}{m_b}
\left(\delta Z_{b,L}^{(1)}
\right.\right.
\nonumber\\
&&{}+\left.
\left.\delta Z_{b,R}^{(1)}\right)
-\frac{1}{8}\left(\delta Z_{b,L}^{(1)}-\delta Z_{b,R}^{(1)}\right)^2
\vphantom{\frac{\delta m_b^{(1)}}{m_b}}\right],
\end{eqnarray}
where
\begin{eqnarray}
\delta_u^{(1)}&=&\frac{1}{2}\delta Z_H^{(1)}
-\frac{1}{2}\,\frac{\delta M_W^{2(1)}}{M_W^2},
\nonumber\\
\delta_u^{(2)}&=&\frac{1}{2}\delta Z_H^{(2)}
-\frac{1}{2}\,\frac{\delta M_W^{2(2)}}{M_W^2} 
+\delta_u^{(1)}\left[-\frac{1}{2}\delta_u^{(1)}
+2(1-\epsilon)
\vphantom{\frac{\delta M_W^{2(1)}}{M_W^2}}\right.
\nonumber\\
&&{}\times\left.\frac{\delta m_t^{(1)}}{m_t}
-2\frac{\delta M_W^{2(1)}}{M_W^2}\right]
-\frac{1}{2}\left(\frac{\delta M_W^{2(1)}}{M_W^2}\right)^2
\end{eqnarray}
are the universal corrections and
\begin{equation}
X^{(i)}=\frac{\delta m_b^{(i)}}{m_b}+\frac{1}{2}\left(\delta Z_{b,L}^{(i)}
+\delta Z_{b,R}^{(i)}\right).
\end{equation}

\begin{figure}[ht]
\begin{center}
\includegraphics[width=0.49\textwidth]{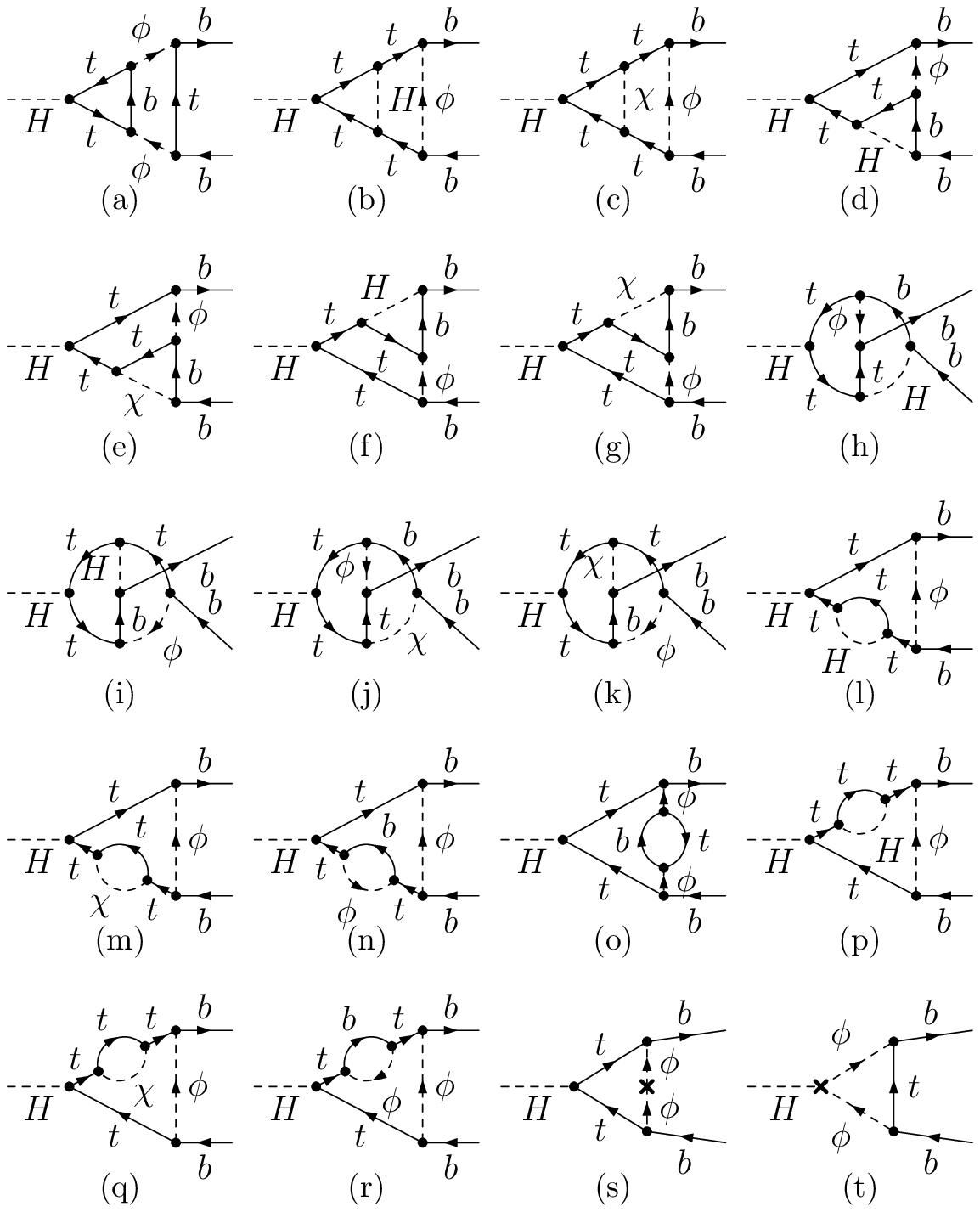}
\end{center}
\caption{\label{DiaHbb2loop}Diagrams contributing to $H\to b\overline{b}$ at
$\mathcal{O}(G_F^2 m_t^4)$.}
\end{figure}

The Feynman diagrams contributing to $A_0^{(2)}$ and $B_0^{(2)}$ are depicted
in Fig.~\ref{DiaHbb2loop}.
They are generated and drawn using the program \texttt{FeynArts}
\cite{Hahn:2000kx} and evaluated using the program \texttt{MATAD}
\cite{MATAD}, which is written in the programming language \texttt{FORM}
\cite{FORM}, by applying the asymptotic-expansion technique (for
a careful introduction, see Ref.~\cite{Smirnov}).
Here, $\chi$ and $\phi$ denote the neutral and charged Higgs-Kibble ghosts
with masses $M_Z$ and $M_W$, respectively. 
The crosses in Figs.~\ref{DiaHbb2loop}(s) and (t) indicate the insertions of
the Higgs-boson mass and tadpole counterterms $i\delta t/v_0$ and 
$-i\left(\delta t/v_0+\delta M_H^2\right)/v_0$ in a $\phi$-boson line and a
$H\phi\phi$ vertex, respectively.
In the soft-Higgs limit, $M_H\ll m_t$, which is underlying our analysis, the
diagrams in Figs.~\ref{DiaHbb2loop}(a)--(s) can also be evaluated by applying
a low-energy theorem (see Ref.~\cite{let} and references cited therein) to the
corresponding $b$-quark self-energy diagrams that emerge by removing the
external Higgs-boson line.
This provides a powerful check for our calculation.
Apart from the diagrams in Fig.~\ref{DiaHbb2loop}, we also need to calculate
the relevant one-particle-irreducible $H$, $b$, and $W$ self-energy diagrams
at two loops.
Furthermore, we need to expand all the relevant one-loop diagrams through
$\mathcal{O}(\epsilon)$.

We are now in a position to present our final results for the universal
correction parameter $\delta_u$ and the relative correction to $\Gamma_b$.
They read
\begin{eqnarray}
\lefteqn{\delta_u=x_tN_c\frac{7}{6}
+x_t^2N_c\left(\frac{29}{2}-6\zeta(2)+N_c\frac{49}{24}\right)}
\nonumber\\
&&{}+x_t\frac{\alpha_s}{\pi}C_FN_c
\left(\frac{19}{12}-\frac{\zeta(2)}{2}\right),
\label{eq:del}\\
\lefteqn{\frac{\Gamma_b}{\Gamma_b^{(0)}}=x_t\left(-6+N_c\frac{7}{3}\right)
+x_t^2\left[-20+N_c(29-12\zeta(2))
\vphantom{\frac{49}{9}}\right.}\nonumber\\
&&{}+\left.N_c^2\frac{49}{9}\right]
+x_t\frac{\alpha_s}{\pi}C_F
\left[-36+N_c\left(\frac{157}{12}-\zeta(2)\right)\right],
\qquad\label{eq:gam}
\end{eqnarray}
where $N_c=3$ and $C_F=(N_c^2-1)/(2N_c)=4/3$ are color factors,
$x_t=(G_Fm_t^2)/(8\pi^2\sqrt{2})$, $\zeta(2)=\pi^2/6$, and
\begin{equation}
\Gamma_b^{(0)}=\frac{\sqrt{2}N_cG_FM_Hm_b^2}{8\pi}
\left(1-\frac{4m_b^2}{M_H^2}\right)^{3/2}.
\label{eq:tree}
\end{equation}
If we convert Eq.~(\ref{eq:del}) to a mixed renormalization scheme which uses
the on-shell definitions for the particle masses and the definitions of the
modified minimal-subtraction ($\overline{\mathrm{MS}}$) scheme for all other
basic parameters, then we find agreement with Eq.~(15) for $x=0$ in the
paper by Djouadi et al.\ \cite{Djouadi:1997rj}.
However, the corresponding result for the electroweak on-shell scheme
presented in their Eq.~(27) for $x=0$ disagrees with our Eq.~(\ref{eq:del}).
We can trace this discrepancy to the absence in their Eq.~(25) of the
additional finite term $\hat{\delta}_{u}^{(1)}\Delta\rho^{(1)}$ which arises
from the renormalization of the one-loop result in their Eq.~(7) according to
the prescription in their Eq.~(18).
The $\mathcal{O}(G_F^2m_t^4)$ term in Eq.~(\ref{eq:gam}) represents a new
result.

In Eqs.~(\ref{eq:del}) and (\ref{eq:gam}), we have also included the two-loop
$\mathcal{O}(\alpha_sG_Fm_t^2)$ corrections
\cite{Kniehl:1994ph,Kniehl:1994ju}, which we reproduced using our
calculational techniques.
As for the QCD renormalization, it is understood that $m_b$ appearing in
Eq.~(\ref{eq:tree}) is defined in the $\overline{\mathrm{MS}}$ scheme as
$m_b=\overline{m}_b(M_H)$, while the electroweak part of the
renormalization remains in the on-shell scheme. 
This modification ensures that large logarithms of the type
$\ln\left(M_H^2/m_b^2\right)$ that would otherwise appear already at
$\mathcal{O}(\alpha_s)$ and spoil the convergence behavior of the
perturbation expansion are properly resummed according to the
renormalization group (RG) \cite{Braaten:1980yq}.
Since we wish to treat $m_t$ on the same footing as $m_b$, we adopt this mixed
scheme for $m_t$ as well.
The analysis at ${\cal O}(\alpha_s^2G_Fm_t^2)$
\cite{Kniehl:1995br,Chetyrkin:1996ke} reveals that Eqs.~(\ref{eq:del}) and
(\ref{eq:gam}) may be further RG-improved by taking $m_t$ and $\alpha_s$ to be
$m_t=\overline{m}_t(m_t)$ and $\alpha_s=\alpha_s^{(6)}(m_t)$, respectively.

Finally, we explore the phenomenological implications of our results.
Adopting from Ref.~\cite{pdg} the values $G_F=1.16637\times10^{-5}$~GeV$^{-2}$,
$\alpha_s^{(5)}(M_Z)=0.1176$, $M_Z=91.1876$~GeV, and
$m_t^\mathrm{pole}=174.2$~GeV for our input parameters, so that
$\alpha_s^{(6)}(m_t)=0.1076$ and $m_t=166.2$~GeV, we evaluate
Eqs.~(\ref{eq:del}) and (\ref{eq:gam}) to $\mathcal{O}(G_Fm_t^2)$,
$\mathcal{O}(G_F^2m_t^4)$, and $\mathcal{O}(\alpha_sG_Fm_t^2)$.
For comparison, we also evaluate the relative corrections to $\Gamma_l$ and
$\Gamma_q$, where $l=e,\mu,\tau$ and $q=u,d,s,c$, which, to the orders
considered here, are given by
\begin{eqnarray}
\frac{\Gamma_l}{\Gamma_l^{(0)}}&=&(1+\delta_u)^2-1,
\\
\frac{\Gamma_q}{\Gamma_q^{(0)}}&=&(1+\Delta_\mathrm{QCD})(1+\delta_u)^2-1,
\label{eq:q}
\end{eqnarray}
where \cite{Braaten:1980yq}
\begin{equation}
\Delta_\mathrm{QCD}=\frac{\alpha_s}{\pi}C_F\frac{17}{4}
\end{equation}
is the ${\cal O}(\alpha_s)$ correction in the limit $m_q\ll M_H$.

The results are listed in Table~\ref{tab:num}.
We observe that the ${\cal O}(G_F^2m_t^4)$ correction to $\Gamma_b$ increases
the enhancement due to the ${\cal O}(G_F m_t^2)$ one by about 16\% and has
more than twice the magnitude of the negative ${\cal O}(\alpha_sG_Fm_t^2)$
one.
Also in the case of $\Gamma_l$, the ${\cal O}(G_F^2m_t^4)$ correction exceeds
the ${\cal O}(\alpha_sG_Fm_t^2)$ one.
The situation is quite different for the case of $\Gamma_q$, which is due to
the additional appearance of the sizeable product term
$2\Delta_\mathrm{QCD}\delta_u^{(1)}$ in Eq.~(\ref{eq:q}).
\begin{table}[t]
\begin{center}
\caption{\label{tab:num}Relative corrections to $\Gamma_\tau$, $\Gamma_c$, and
$\Gamma_b$ at $\mathcal{O}(G_Fm_t^2)$, $\mathcal{O}(G_F^2m_t^4)$, and
$\mathcal{O}(\alpha_sG_Fm_t^2)$.}
\begin{tabular}{|c|ccc|}
\hline
Order & $\Gamma_\tau/\Gamma_\tau^{(0)}$ & $\Gamma_c/\Gamma_c^{(0)}$ &
$\Gamma_b/\Gamma_b^{(0)}$ \\
\hline
$\mathcal{O}(G_Fm_t^2)$ & $+2.021\%$ & $+2.021\%$ & $+0.289\%$ \\
$\mathcal{O}(G_F^2m_t^4)$ & $+0.064\%$ & $+0.064\%$ & $+0.047\%$ \\
$\mathcal{O}(\alpha_sG_Fm_t^2)$ & $+0.060\%$ & $+0.452\%$ & $-0.022\%$ \\
\hline
\end{tabular}
\end{center}
\end{table}

In conclusion, we analytically calculated the dominant electroweak two-loop
correction, of order ${\cal O}(G_F^2m_t^4)$, to the $H\to b\overline{b}$ decay
width $\Gamma_b$ of an intermediate-mass Higgs boson, with $M_H\ll m_t$.
We performed various checks for our analysis.
The ultraviolet divergences cancelled through genuine two-loop renormalization.
Our final result is devoid of infrared divergences related to infinitesimal
scalar-boson masses.
We reproduced those $Hb\overline{b}$ vertex diagrams where the external Higgs
boson is coupled to an internal top-quark line, which we had computed 
directly, through application of a low-energy theorem.
After switching to a hybrid renormalization scheme, our ${\cal O}(G_F^2m_t^4)$
result for the universal correction $\delta_u$ agrees with
Ref.~\cite{Djouadi:1997rj}.
Using our computational techniques, we also recovered the
${\cal O}(\alpha_sG_Fm_t^2)$ corrections to $\delta_u$ and $\Gamma_b$.
The ${\cal O}(G_F^2m_t^4)$ correction to $\Gamma_b$ amplifies the familiar
enhancement due to the ${\cal O}(G_Fm_t^2)$ correction by about $+16\%$ and
thus more than compensates the screening by about $-8\%$ through QCD effects
of ${\cal O}(\alpha_sG_Fm_t^2)$.

We like to thank Paolo Gambino and Matthias Steinhauser for fruitful
discussions.
This work was supported in part by the German Federal Ministry for Education
and Research BMBF through Grant No.\ 05~HT6GUA and by the German Research
Foundation DFG through Graduate School No.\ GRK~602 {\it Future
Developments in Particle Physics}.

\end{document}